%% file: main.tex
\documentclass[reprint]{revtex4-2}

\usepackage{graphicx}
\usepackage{dcolumn}
\usepackage{bm}
\usepackage[hidelinks]{hyperref}

\begin{document}

\preprint{APS/123-QED}

\title{Electrostatics overcome acoustic collapse to\\assemble, adapt, and activate levitated matter}

\author{Sue Shi}
 \email{sue.shi@ist.ac.at}
\author{Maximilian C.~H\"ubl}%

\author{Galien P.~Grosjean}

\author{Carl P.~Goodrich}

\author{Scott R.~Waitukaitis}
 \email{scott.waitukaitis@ist.ac.at}

\affiliation{%
Institute of Science and Technology Austria (ISTA),\\
  Am Campus 1, 3400 Klosterneuburg, Austria
}%


\begin{abstract}
Acoustic levitation provides a unique method for manipulating small particles as it completely evades effects from gravity, container walls, or physical handling. These advantages make it a tantalizing platform for studying complex phenomena in many-particle systems, save for one severe limitation---particles suspended by sound interact via acoustic scattering forces, which cause them to merge into a single dense object. To overcome this ``acoustic collapse'', we have developed a strategy that combines acoustic levitation with controlled electrostatic charging to assemble, adapt, and activate complex, many-particle systems. The key idea is to introduce electrostatic repulsion, which renders a so-called ``mermaid'' potential where interactions are attractive at short range and repulsive at long range. By controlling the balance between attraction/repulsion, we are able to levitate fully expanded structures where all particles are separated, fully collapsed structures where they are all in contact, and hybrid ones consisting of both expanded and collapsed components. We find that fully collapsed and expanded structures are inherently stable, whereas hybrid ones exhibit transient stability governed by acoustically unstable dimers. Furthermore, we show how electrostatics allow us to adapt between configurations on the fly, either by quasistatic discharge or discrete up/down charge steps.  Finally,  we demonstrate how large structures experience selective energy pumping from the acoustic field---thrusting some particles into motion while others remain stationary---leading to complex dynamics including coupled rotations and oscillations. Our approach provides an easy-to-implement and easy-to-understand solution to the pervasive problem of acoustic collapse, while simultaneously providing new insights into the assembly and activation of many-particle systems with complex interactions.  
\end{abstract}

\maketitle


\section{\label{sec:level1}Introduction}

Acoustic levitation enables objects in the size range from tens of microns to millimeters to be suspended in air by sound \cite{santesson2004airborne, vandaele2005non, foresti2013acoustophoretic, marzo2015holographic, ahmed2016rotational, llewellyn20163d, melde2016holograms, memoli2017metamaterial, shi2019general, lim2019cluster, andrade2020acoustic}. In the most common approach, an acoustic ``trap'' consists of a standing wave created between emitter(s) and reflector(s), where particles are drawn to pressure nodes (or, if they are less dense than the medium, antinodes) \cite{andrade2018review}. A particle in a trap is lifted and confined by the primary acoustic force, which arises from the momentum transferred by the incident field to the particle as sound scatters off its surface \cite{king1934acoustic, yosioka1955acoustic, gor1962forces, bruus2012acoustofluidics}. However, this is not the only force at play---the sound scattered off a given particle modifies the acoustic field at all other particles, leading to interparticle forces that in most situations are attractive \cite{konig1891hydrodynamisch, bjerknes1906fields, embleton1962mutual,crum1975bjerknes,zhuk1985hydrodynamic,nyborg1989theoretical,doinikov1995mutual,zheng1995acoustic,doinikov1996interaction,doinikov1996mutual,doinikov1999bjerknes,doinikov2001acoustic,doinikov2002viscous,pelekasis2004secondary,silva2014ideal}. As a result of these scattering forces, levitated many-particle systems suffer from ``acoustic collapse'', where all particles are drawn into physical contact in a single close-packed clump \cite{lim2019cluster, lim2022mechanical}. This is a severe limitation.  If it could be avoided, \textit{i.e.}, if there were a method to levitate many particles and overcome scattering forces to keep them physically separated, then the usefulness of acoustic levitation would be significantly extended. For instance, maintaining controlled separation between levitated droplets could enable precise microscale chemical reactions and synthesis within acoustic traps \cite{dittrich2006lab, teh2008droplet, evander2012acoustofluidics}, while doing the same with solid particles could be useful in levitation-based approaches to 3D printing \cite{ngo2018additive} or in studies aimed at understanding how particles aggregate due to long range interactions, \textit{e.g.},~in planet formation \cite{onyeagusi2022measuring}. More fundamentally, overcoming collapse would enable acoustic levitation to be used as a platform to study complex interparticle forces---\textit{e.g.}, the acoustic scattering force itself, which theory indicates is both non-reciprocal \cite{King.2025} and many-body \cite{st2023dynamics}. Such interactions are of particular interest because they can give rise to novel collective behaviors and transport phenomena, relevant to emerging fields like active matter and topological materials. \par

Recent works have mitigated the worst effects of collapse in two ways.  In the first, the strategy has been to work with extremely fine powders (\textit{i.e.}, particles limited to be less than tens of microns in size), where repulsive acoustic streaming forces become significant and give particles a bit of separation \cite{wu2023hydrodynamic, wu2025pattern}. This vein has shed many insights into novel physics caused by the streaming force, but is limited to small objects and only gives very small ($\sim$10 \textmu{m}) separations. Moreover, acoustic streaming forces, like acoustic scattering forces, are created by the acoustic field; hence, they are not easily tuned independently. A second approach that counters collapse has been to use acoustic holography, which, in essence, ``sculpts'' the sound field so that there are multiple nodes \cite{melde2016holograms, marzo2019holographic, melde2023compact, shi2025acoustic}. This method enables exquisite patterning of the locations where particles get trapped, and beyond this allows targeted injection of energy. However, it neither prevents collapse from occurring within a given node---as attractive scattering forces still prevail in the particle-particle interaction---nor permits dynamic reconfiguration during experiments.\par

In this work, we introduce a new strategy to overcome acoustic collapse. The key idea is to use controlled electrostatic repulsion to counter scattering forces and keep particles separated. Electrostatic interactions constitute a degree of freedom that is completely independent from the acoustic field, hence opening up the possibility to explore fully collapsed configurations, perfectly expanded ones, and complex hybrids in between. Our method necessitates conductive media, but beyond this, suffers no particular restrictions concerning particle size or number. With two methods to change the charge of particles \textit{in situ}, we show how electrostatics allow us to adapt the system on the fly. Finally, we demonstrate and explore how larger configurations consisting of both collapsed and expanded components selectively harvest energy from the incident field, leading to complex rotational and oscillatory dynamics. Our approach thus paves the way toward deeper insights into non-equilibrium assembly and control in acoustically driven systems, significantly expanding the toolkit for studying complex, many-particle interactions and dynamics.

\begin{figure*}[t]
\centering
\includegraphics[width=\linewidth]{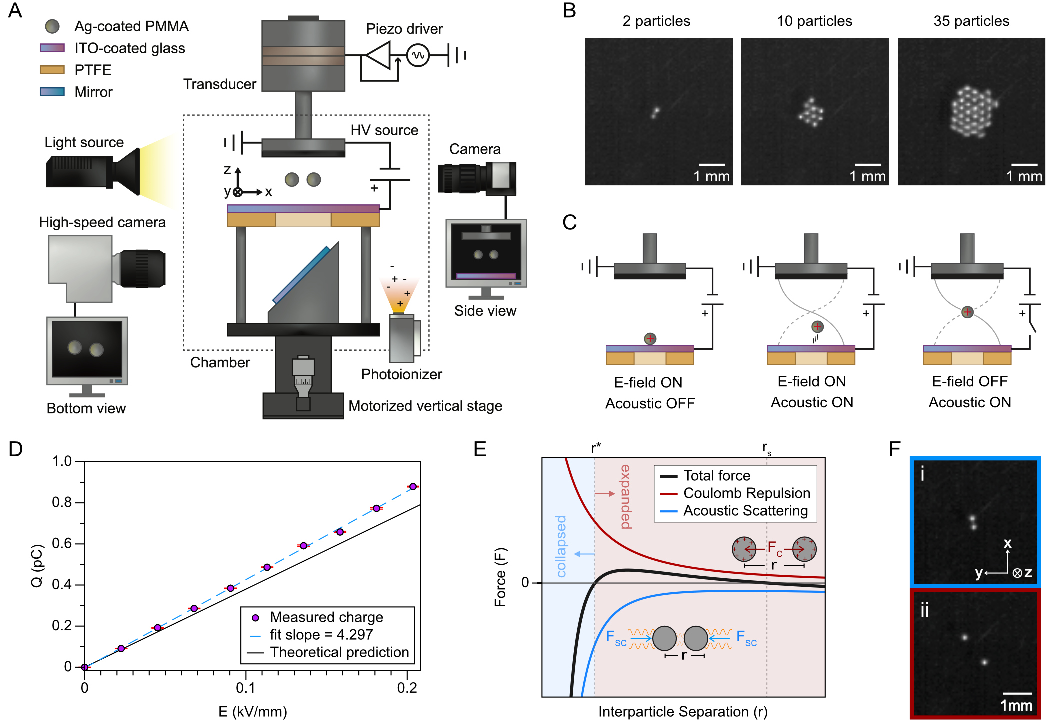}
\caption{Acoustic levitation of like-charged particles. (A) Schematic of the experimental setup. A transducer generates a standing wave between its horn and a reflecting plate. The plate is composed of ITO-coated glass layered with PTFE for electrical insulation and mounted on a motorized vertical translation stage for precise alignment of the acoustic cavity. Conductive particles placed on the plate acquire charge when the ITO surface is connected to a high-voltage source. Charged particles levitate in the acoustic cavity and can be imaged from both the side and bottom via a mirror. A soft X-ray source enables controlled discharge. (B) Bottom-view images of compact rafts consisting of 2, 10, and 35 particles. (C) Schematics of experimental procedures: initially, particles acquire charge via contact with the biased ITO plate; next, we switch on the acoustic field to levitate charged particles; finally, the ITO plate is no longer voltage biased and charged particles levitate only in the acoustic field. (D) Measured charge $Q$ on a silver-coated PMMA particle (diameter $\mathrm{d \simeq 288~\mu m}$) as a function of the applied electric field $\mathrm{E}$. Purple circles represent the mean charge measured among (at most) 60 trials per data point. The dashed blue line is a linear fit to the data points, with a slope of $\sim$4.3. The black line shows the theoretical prediction based on Maxwell's approximation for a conducting sphere on a conducting plane. Error bars represent the standard deviation. (E) Interparticle forces as a function of center-to-center separation $r$. The repulsive Coulomb force (red) and attractive acoustic scattering force (blue) are shown along with their sum (black), representing the total interaction force. The total force crosses zero at $r = r_*$ and $r = r_s$, corresponding to an unstable and a stable equilibrium point, respectively. (F) Bottom-view images of two levitated particles: (i) in contact when acoustic scattering forces dominate, and (ii) separated when Coulomb repulsion dominates.}
\label{fig1}
\end{figure*}

\section{Results}

The experimental setup is illustrated in Fig.~\ref{fig1}\textit{A}; while we succinctly explain its operation here, full details are provided in the \textit{Methods} section. We drive a piezoelectric transducer near its electromechanical resonance to generate a standing wave of sound between its emitting horn and a reflecting glass plate below. Particles levitate in the node of this standing wave, where we observe their motion with cameras from both the side and bottom. The reflector plate is coated with a thin layer of indium tin oxide (ITO) and mounted on a motorized vertical translation stage for precise detuning of the cavity gap, $h$. The chamber houses a photoionizer, which allows us to discharge all objects inside. Unless otherwise specified, the particles we work with are silver-coated poly(methyl methacrylate) (PMMA) microspheres with diameters between 250 and 300~\textmu{m}. Without any special thought given to the charge of particles, \textit{i.e.}~if we simply turn on the system with some particles present, they levitate and collapse into dense rafts at the trap center due to the attractive acoustic scattering forces, as seen in most acoustic levitation experiments \cite{lim2019cluster, lim2022mechanical, brown2024direct} and shown in Fig.~\ref{fig1}\textit{B}.

\subsection*{Acoustically Levitating Charged Mermaid Particles}

Figure \ref{fig1}\textit{C} shows how we imbue the particles with a precisely controllable amount of electrical charge to avoid collapse. In the simplest implementation, we let a particle rest on the ITO plate with the acoustic field off, but with a high-voltage DC potential, $V$, applied between the plate and transducer. This leads to a capacitive buildup of charge on the particle, $Q$, which can be estimated from Maxwell's solution \cite{maxwell1892treatise, fish1967conductive} for two contacting spheres in the limit that one (\textit{i.e.}, the lower plate) has infinite radius, 
\begin{equation}
\label{chargeonsphere}
Q \approx \frac{2}{3}\frac{V}{h}\pi^{3}a^{2}\epsilon_{\mathrm{0}}.
\end{equation}
Here, $a$ is the radius of the particle and $\epsilon_{\mathrm{0}}$ is the vacuum permittivity. Next, we switch on the acoustic field and quickly ({$\sim$10~ms}) thereafter switch off the electric field. At the instant when both fields are on, and depending on the electric field strength (see \textit{SI Appendix}), either can be sufficient to launch the particle toward the trap center, charge in tow. Within a few seconds, the initial kinetic energy from the launch is viscously dissipated with the air, and the particle stably levitates in the trap, in principle with the charge given by Eq.~\ref{chargeonsphere}.  

To confirm this approach quantitatively, we work with a single particle and measure the charge it acquires, $Q$, as a function of the electric field strength, $E=V/h$. For technical reasons, these measurements are performed in a separate but nearly identical levitation setup; see \textit{SI Appendix}. In order to perform multiple measurements with the same particle, we slightly alter the protocol shown in Fig.~\ref{fig1}\textit{C} by:~(1) starting with the particle levitated, (2) turning off the acoustic field to let it fall, (3) simultaneously turning on the electric field so it acquires charge when it contacts (bounces off) the substrate, and (4) then reinitiating the acoustic field (and turning off the electric field) at just the right moment to ``catch'' it again in the trap. Once the particle is stably levitated, we measure its dynamical response to an AC electric field to obtain its charge.  The results shown in Fig.~\ref{fig1}\textit{D} are in good agreement with the theoretical prediction given by Eq.~\ref{chargeonsphere}, with no adjustable parameters. Details on calibration for the main setup can be found in the \textit{SI Appendix}.

Our charging method works equally well for multiple particles in the trap, allowing us to efficiently load particles with virtually any charge we want, only limited by the dielectric breakdown of air. What does the presence of charge imply for interparticle forces?  Fig.~\ref{fig1}\textit{E} conceptually depicts the situation for a pair of particles. Both particles experience the aforementioned acoustic scattering force, which, in the ``Rayleigh limit'', \textit{i.e.}, if the particles both are spheres with radius $a$, and if $a \ll r \ll \lambda$ ($r$ is the interparticle spacing, and $\lambda$ the acoustic wavelength) \cite{wu2023hydrodynamic}, scales as: 
\begin{equation}
\label{scatter}
F_{\mathrm{sc}} \propto - \frac{a^{6} E_{ac}}{r^4},
\end{equation}
where $E_{ac}$ is the acoustic energy density of the incident field. On the other hand, the repulsive force for two like-charged spheres is given by Coulomb’s law, 
\begin{equation}
\label{CoulombLaw}
F_{\mathrm{C}} = \frac{Q^2}{4\pi\epsilon_{0}r^2}.
\end{equation}
 Due to the fact that $F_{\mathrm{sc}}$ decays as $r^{-4}$ while $F_{\mathrm{C}}$ decays as $r^{-2}$, the net interaction is attractive at short separations but repulsive at longer distances, resulting in a so-called ``mermaid'' potential. If we additionally consider the primary acoustic force that draws particles to the trap center, we see that there are two predicted configurations for a pair of particles: collapsed or expanded. We confirm this prediction experimentally in Fig.~\ref{fig1}\textit{F}, where we put two particles on the ITO plate, charge them, and load them into the trap. After about a  1~s transient during which the air dampens out the initial kinetic energy from the launch, the particles settle into one of the two predicted configurations (see Movies S1 and S2). Which structure forms depends on the particle trajectories during the transient stage: if the particles happen to approach too closely, the short-range acoustic scattering force dominates, resulting in the collapsed final state; conversely, if the particles maintain sufficient separation or have enough kinetic energy to overcome acoustic attraction, then Coulomb repulsion stabilizes them into the expanded equilibrium. The statistics of structure formation will be discussed in detail in the next section. 

\begin{figure*}[t!]
\centering
\includegraphics[width=178mm, height=111mm]{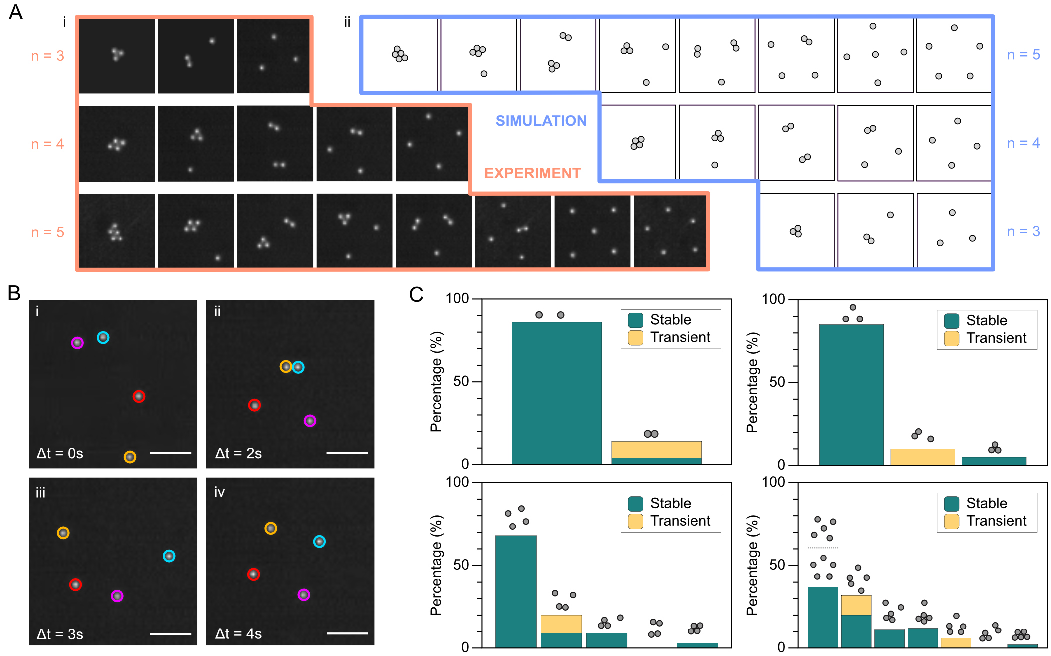}
\caption{Structure formation and statistics for $\mathrm{n = 2-5}$ particles. (A) Panel (i) shows bottom-view images of self-assembled clusters of $\mathrm{n =3}$, 4, and 5 levitated charged particles (diameter $=250-300~\mathrm{\mu m}$). Beside fully collapsed configurations where no particle is separated and fully expanded ones where all particles are separated, we observe a variety of hybrid configurations, which consist of mixed arrangements between collapsed and expanded clusters. The corresponding simulated structures shown in panel (ii) are obtained via Molecular Dynamics and fully reproduce the experimental results. (B) Bottom-view image sequence showing the transient formation and rearrangement of a hybrid structure with an unstable dimer. Colored circles identify the same particle through all four panels. Scales bars are 1~mm. (C) Experimental bar plots showing the occurrence frequency and stability of each configuration, obtained by 400~V ITO bias. Stable occurrences (teal) persist for more than 15 seconds, while transient occurrences (yellow) rearrange or disintegrate within 15 seconds.}
\label{fig2}
\end{figure*}

\subsection*{Configurations and Stability at Low Particle Numbers }

Building on the elementary example of two particles, we next examine how competing acoustic and Coulomb forces affect the assembly of systems consisting  of up to $\mathrm{n=5}$ particles. While two like‐charged particles can form either a collapsed dimer or a fully expanded pair, moving to three or beyond introduces the possibility for hybrids to form---\textit{i.e.}, configurations with both collapsed and expanded components. For instance, the middle image of the top row in Fig.~\ref{fig2}\textit{A}~\textit{i} shows an $\mathrm{n=3}$ hybrid consisting of a collapsed dimer plus one separated particle. Such a configuration arises when, after initial kinetic energy from the launch is sufficiently damped out, two of the three particles are separated by less than the critical distance while one remains repelled. For $\mathrm{n=4}$, there are three distinct hybrid configurations, and for $\mathrm{n=5}$, there are five. Up to $\mathrm{n=5}$, only one compact configuration exists for each cluster size: a triangle for $\mathrm{n=3}$, a rhombus for $\mathrm{n=4}$, and a trapezoid for $\mathrm{n=5}$. Starting at $\mathrm{n=6}$, the system begins to have multiple, geometrically distinct compact configurations \cite{lim2019cluster}. Notably, while $\mathrm{n=3}$ and $\mathrm{n=4}$ each yield only one fully expanded configuration, $\mathrm{n=5}$ exhibits  two distinct possibilities. \par

With caveats, we qualitatively reproduce most aspects of the observed $\mathrm{n>2}$ configurations by assuming the applicability of pairwise scattering forces and Coulomb's law across all particle pairs. We ascertain this by finding stable configurations in molecular dynamics (MD) simulations that incorporate acoustic forces calculated from second-order perturbation theory in the Rayleigh limit~\cite{bruus2012acoustofluidics, silva2014ideal}, soft-sphere repulsion to prevent overlap, and Stokes drag (see \textit{Methods} for full details). The results are shown in Fig.~\ref{fig2}\textit{A~ii}, with four additional ``chain'' configurations that are absent in experiments shown in Fig.~S1. Aside from these exceptions, the simulated configurations shown in Fig.~\ref{fig2}\textit{A~ii} closely match the experimentally observed ones in Fig.~\ref{fig2}\textit{A~i}. The slight deviations in the experimental results (\textit{e.g.}, the unequal side lengths of the expanded pentagon for $\mathrm{n=5}$) can be explained by as little as 10\% size variability in the particles, which affects both the acoustic forces and the acquired charge (see Fig.~S2). As we discuss in the \textit{SI Appendix}, the intrinsic non-reciprocal \cite{King.2025} and many-body \cite{st2023dynamics, lim2022mechanical} nature of acoustic scattering forces means that applying our pairwise model across all interactions is strictly incorrect, even if it results in the same configurations observed in experiments. In particular, Ref.~\cite{King.2025} has shown that the lowest-order results in the Rayleigh limit fail to account for small non-reciprocal components that are present whenever
particles are different,\textit{e.g.} in size or shape. \par

A manifestation of these exotic aspects of the scattering force may be present in the ``long-term'' stability of configurations. Both fully collapsed and fully expanded configurations at small n remain stable for hours or longer after formation. In contrast, hybrids that involve a dimer frequently exhibit only transient stability, where they eventually separate and reconfigure into a different motif. We highlight an example of this in Fig.~\ref{fig2}\textit{B} and Movie S3. Here, four particles initially at rest on the ITO plate (panel~i) assemble into a hybrid configuration with a collapsed dimer and two monomers (panel~ii) within the first two seconds. By $\mathrm{\Delta t=3~s}$, the dimer spontaneously begins to oscillate, ultimately destabilizing and separating (panel~iii).  Subsequently, the entire system rearranges and settles into a stable, expanded-square configuration (panel~iv). A comprehensive understanding of this instability is beyond the scope of this study, but we have determined that it originates from acoustic rather than electrostatic forces. To rule out an electrostatic origin, we routinely observe that uncharged dimers exhibit spontaneous and sporadic oscillations, which occasionally amplify over time leading to their separation (see Movies S4 and S5). Similar instabilities have been reported in single-mode acoustic levitators and attributed to resonant frequency shifts within the cavity due to the presence of the objects \cite{rudnick1990oscillational, putterman1989acoustic}. In light of the results of Ref.~\cite{King.2025}, a plausible alternative is that it could be a signature of non-reciprocity. From a sound-scattering perspective, collapsed objects (\textit{e.g.}, dimers) behave like non-spherical single particles. This means that their acoustic interactions with other scatterers in the cavity (single particles, other collapsed objects, or even the emitter/reflector) are generally non-conservative and highly sensitive to orientation. Qualitatively, this might explain how the oscillations that preempt destabilization develop---non-conservative interactions with other particles or perhaps even the boundaries pump in energy.  We speculate that such instabilities are suppressed in higher $\mathrm{n}$ collapsed objects due to the relatively less pronounced asymmetries in their shapes and the increased energy barrier associated with breaking additional acoustic bonds. \par

In Fig.~\ref{fig2}\textit{C}, we quantify the frequency and stability of every configuration observed at 400~V ITO bias, which is equivalent to $\sim 0.5~\mathrm{pC}$ of charge per particle. We perform an exhaustive search (several hundred trials) for $\mathrm{n=2}$ to 5 and record what configuration forms in the trap and whether or not it persists for longer than 15~s (classified as stable) or reconfigures within this interval (transient).  At these particular acoustic/electrostatic strengths, the statistics confirm that fully expanded clusters are reliably stabilized by Coulomb repulsion, collapsed clusters by multiple acoustic bonds, and hybrids only when they do not incorporate an inherently unstable  dimer. Additionally, Fig.~\ref{fig2}\textit{C} illustrates that fully expanded configurations dominate at small particle numbers, but this changes as $\mathrm{n}$ grows. This seems to be a numbers game related to the fast-growing multiplicity of possible collapsed arrangements and the decreasing spatial margin to avoid short range attraction as $\mathrm{n}$ grows. Indeed, at high $\mathrm{n}$ we find that the vast majority of configurations are hybrids with a single large collapsed component (see Fig.~\ref{fig4}). In the \textit{SI Appendix}, we describe the outcomes of these statistics in more detail, and in particular show how they change if we modulate the balance of acoustics/electrostatics by using a higher ITO bias voltage (Fig.~S3).\par

\subsection*{Electrostatic Adaptation on the Fly}

\begin{figure}
\centering
\includegraphics[width=\linewidth]{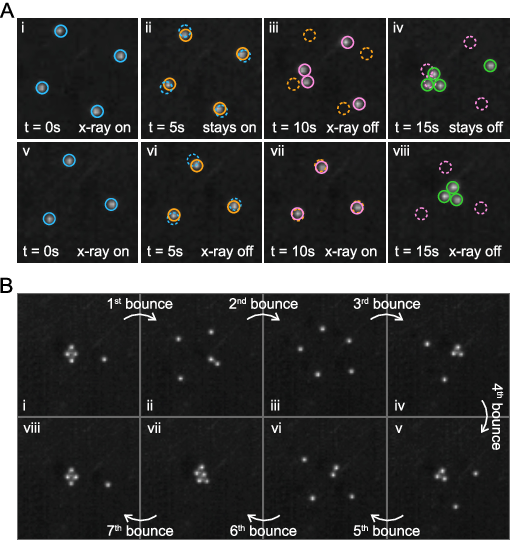}
\caption{Manipulating assembled configurations via controlled discharges and bounce-charging. (A) Bottom-view images showing structural evolution under two discharge protocols. Panels (i–iv) show a single, terminated discharge event. Panels (v–viii) illustrate a two-step discharge sequence using timed X-ray exposure.. Solid circles indicate current particle positions; dashed circles indicate positions from the previous frame. (B) Bottom-view sequence showing configuration shuffling with the bounce-charging method. Each panel corresponds to a successive bounce on the ITO plate biased at $\mathrm{500~V}$, leading to structural rearrangements.}
\label{fig3}
\end{figure}

As we have shown, our approach of using repulsive electrostatics allows us to overcome scattering forces and avoid acoustic collapse, resulting in multifarious assembled configurations due to the complex potential landscape between our particles. We now demonstrate how it also allows us to adapt between such configurations, as achieved by  two complementary methods: controlled discharge and ``bounce charging''. The first of these  is demonstrated in Fig.~\ref{fig3}\textit{A}. We use the photoionizer to increase the conductivity of the air in the chamber, which causes the particles to quasistatically discharge as they draw in ions of the opposite sign.  As their charges diminish, Coulomb repulsion weakens, causing the particles to move closer together. By additionally controlling the durations and intervals of ``pulses'' of the photoionizer, we exert a high degree of control over this process. For instance, panels i-iv show a single-pulse discharge process with an initially expanded four particle configuration (Movie S6). Continuous ionization over 10~s initially draws the entire system inward (panels i-ii, blue to orange outlines). Suddenly, two particles that become sufficiently close to cross their tipping point collapse into an unstable dimer (panels ii-iii, orange to pink outlines). Although the discharge pulse stops at $t=10$~s, the dimer oscillates due to its inherent instability, ultimately leading to the sudden capture of one of the two remaining separated particles, forming a stable collapsed triangle (panels iii-iv, pink to green outlines). The final stable configuration consists of this collapsed triangle and a monomer, remaining intact indefinitely after discharge ends. Fig.~\ref{fig3}\textit{A} panels v-viii depict a two-pulse discharge sequence (Movie S7). Here, an initial 5~s pulse partially discharges the fully expanded triangle, causing a slight inward repositioning of the spheres (panels v-vi,  blue to orange outlines). With ionization paused for the following 5~s, the particles hold their positions stably (panel vii, overlapping orange and pink outlines). Then, with a second subsequent pulse, we further discharge the system, ultimately forcing the expanded configuration into an acoustically collapsed triangle (panels vii-viii, pink to green outlines).\par

Figure~\ref{fig3}\textit{B} demonstrates our second method for \textit{in situ}, electrostatic adaptation---bounce-charging. In essence, this method is nearly identical to the process used to alter the charge of a single particle in the validation of Eq.~\ref{chargeonsphere}.  We (1) start with a particular configuration of $\mathrm{n}$ particles levitated, (2) turn off the acoustic field so they all fall, (3) simultaneously turn on the electric field so particles acquire charge when they contact (bounce off) the bottom plate, and (4) then reinitiate the acoustic field (and turn off the electric field) at just the right moment to ``catch'' all particles again. Importantly, during the particles' contact with the bottom plate, we have the tunability to use whatever voltage we want, short of igniting a spark in the cavity. Fig.~\ref{fig3}\textit{B} shows how this works in practice for the case of five particles with no change to $V$ during bounces. Initially, the particles start in a $4+1$ hybrid (panel~i). The first bounce-charging iteration drives a rearrangement into a $2+1+1+1$ hybrid, the second into an expanded pentagon, and so on, with each successive bounce causing the particles to reshuffle structurally upon reinitiation of levitation (Movies S8 and S9). Although the precise configuration formed after each bounce is not strictly deterministic, we can ``nudge'' the system in a desired direction. For example, if we start with an expanded configuration but we desire a collapsed one, we can use lower voltages. We can also move from collapsed toward expanded states by using high voltages, although as Figs.~\ref{fig2}\textit{C} and S3\textit{C} suggest, this becomes less efficient at higher $\mathrm{n}$. \par

\subsection*{Structure-Coupled Active Dynamics of Large Hybrids}

\begin{figure*}
\centering
\includegraphics[width=\textwidth]{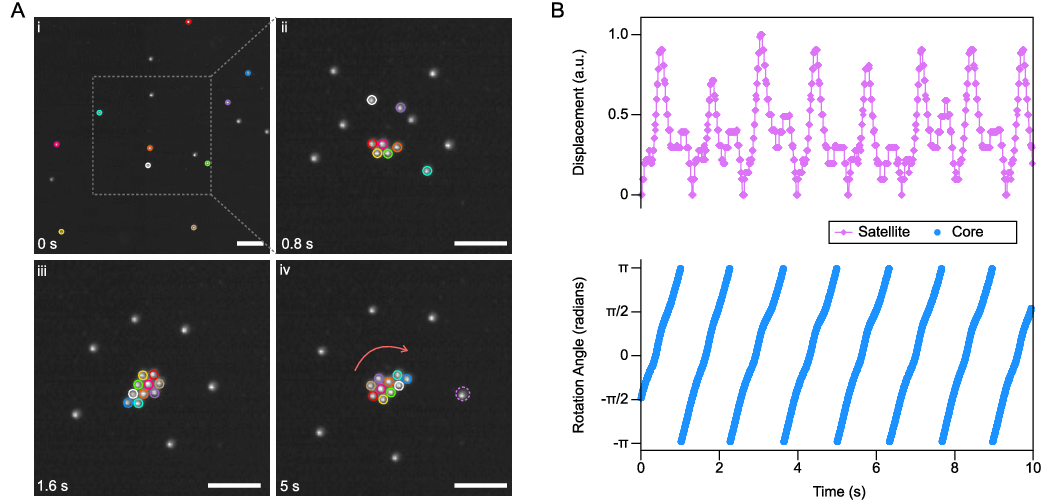}
\caption{Assembly and dynamics of large hybrid configurations. (A) Bottom-view time-lapse of the ``acoustic clock'', assembled from smaller particles with diameters in the range of $180-212~\mathrm{\mu m}$. Colored circles trace the trajectories of individual particles from their initial positions in (i) to the formation of the collapsed central cluster in (iv). The red arrow in (iv) indicates the direction of cluster rotation. Panels ii-iv are $2\times$ zoom views of the area within the dashed square in panel i. All scale bars are 1~mm. (B) Purple curve shows the normalized displacement of a satellite particle (indicated by dashed purple circle in panel iv); blue curve plots the core’s rotation angle versus time, showing periodic coupling between core rotation and satellite particle motion.}
\label{fig4}
\end{figure*}

We now turn our attention to substantially larger systems, which introduce two new features: (1) they assemble into hybrids with collapsed centers surrounded by individual particles, and (2) they harvest energy from the acoustic field to exhibit complex rotational $+$ oscillational dynamics. Spontaneous rotation of levitated objects inside acoustic cavities is observed frequently, where its onset has been attributed to a variety of factors including oscillational instabilities \cite{rudnick1990oscillational}, asymmetries of the trapping field \cite{xiu2017rotation, onyeagusi2022measuring}, acoustic streaming flows \cite{trinh1994experimental}, horn geometry \cite{baer2011analysis}, non-reciprocal forces~\cite{King.2025}, and off-resonance driving \cite{lim2022mechanical}. Our experiments do not shed any light on the origin of rotation, but instead show new features that are unique due to the fact that some particles avoid collapse.\par  

Figure~\ref{fig4}\textit{A} presents an image sequence illustrating the formation of one frequently encountered motif, which, for reasons that will become clear, we call an ``acoustic clock'' (Movie S10). Colored circles highlight the locations of individual particles as they start from rest on the ITO substrate (0~s, panel~i), are lifted up once levitation initiates, and then assemble. A collapsed component develops first, consisting of five particles by 0.8~s (panel ii) and finishing with 10 particles by 1.6~s (panel iii). Subsequently, the remaining particles assemble as ``hour markers'', electrostatically separated by significant distances from the collapsed ``clock hand''. Not long after, the hand spontaneously begins to rotate in a clockwise fashion, as indicated by the red arrow in panel iv.\par

The rotational dynamics we observe in objects like this acoustic clock are unique for several reasons. Foremost, not all particles in the system rotate---only the collapsed component. The hour-marker particles that evaded collapse remain at essentially fixed angular positions around the center. Due to electrostatic interactions, however, they ``tick'', \textit{i.e.}, oscillate radially inward and outward, in response to the rotating hand. This is easily seen qualitatively in the SI video, but we can also see it quantitatively. In Fig.~\ref{fig4}\textit{B}, we plot the angular position of the hand as well as the radial position of one of the hour-marker particles. In response to the hand's rotation, the particle oscillates, with an overall period set by the hand.  In the \textit{SI Appendix}, we show different dynamical motifs, including ``Ferris wheels'' and edge rotators that couple to electrostatic breathing modes (see Movies S11 and S12). We also demonstrate the acoustic strength dependence of one such motif (Movie S13). \par 

\section{Discussion}
We have introduced a strategy to overcome acoustic collapse in sound-levitated systems. Our approach relies on precisely controlling the charge on electrically conductive particles, which allows repulsive Coulomb forces to overcome scattering forces and keep particles separated. From stable, fully expanded structures to fully collapsed structures, and myriad hybrid states in between, we observe complex assembled configurations, whose relative occurrence is tunable via the amount of charge imbued. By altering the charge after particles are levitated via air ionization or bounce-charging, we adapt between configurations. As with many other acoustically levitated systems, we observe the injection of rotational energy into particles, but selectively to collapsed components---those particles that avoid collapse do not rotate, but do move in electrostatic response. Hence, our method allows access to previously restricted particle configurations, the ability to switch between them, and the possibility to selectively activate them---all without complex engineering or reliance on difficult-to-tune forces. This makes our approach broadly accessible and powerful in its simplicity. \par

Avoiding acoustic collapse opens up the possibility to use acoustic levitation to study complex interactions in many-particle systems in gravity- and container-free conditions. The most obvious target in this vein is to study the non-reciprocal and many-body aspects of the acoustic scattering force itself. As we have already intimated, non-reciprocal/many-body effects may be lurking in some phenomena we have presented here, \textit{e.g.},~dimer destabilization and the prevalence of central-core hybrids at large n.  In the \textit{SI Appendix}, we show further examples that suggest the presence of effects, for instance the ``chasing'' of some fully expanded systems (Movie S14). With the potential to work with larger or aspherical particles, we expect our approach to give access to these and other types of complex many-body phenomena.\par

\section{Materials and Methods}
\subsection*{Experiment}
 The ultrasonic standing‐wave acoustic cavity was driven by a bolt-clamped Langevin-type transducer based on a textbook design example \cite{andrade2020design}, with piezoelectric elements from a commercially available transducer (STH-S44-3030-A, Micromechatronics). The transducer was powered by an amplifier with feedback circuitry (PDUS210-FLEX, PiezoDrive), operating at or near its electromechanical resonance around $f_0 \approx 51~\mathrm{kHz}$, resulting in an emitted sound wavelength in air of $\lambda_0 \approx~$6.8~mm. The transducer horn, measuring 20~$\mathrm{mm}$ in diameter and 3~$\mathrm{mm}$ in thickness, featured a flat bottom surface coated with black, non-conducting paint to enhance particle imaging and prevent charge loss upon particle collision. The bottom reflecting plate consisted of indium tin oxide (ITO) coated glass ($50 \times 50 \times 4~\mathrm{mm}$, UQG Optics), supported by a polytetrafluoroethylene (PTFE) holder for electrical insulation. It was mounted onto a motorized vertical translation stage actuated by a DC servo motor (Z925B, Thorlabs) controlled by a brushed DC servo motor controller (KDC101, Thorlabs), enabling micrometer-scale adjustments to the cavity height. The gap between the transducer horn surface and the ITO surface was adjusted to $\lambda_0/2\approx 3.4$~mm. Precise tuning of the cavity resonance height was verified using an optical microphone (Eta100 Ultra, Xarion Laser Acoustics). 

We used conductive silver-coated poly(methyl methacrylate) microspheres (Cospheric) of two different sizes: microspheres with diameters between $250-300$~\textmu{m} and density of $1.2~\mathrm{g/cc}$ were used in most experiments, while smaller microspheres with diameters between $180-212$~\textmu{m} and density of $1.3~\mathrm{g/cc}$ were used in selected experiments, as specified in the main text.
 
 A DC/DC high voltage precision power supply (6LE24-P20 Ultravolt, Advanced Energy) provided a bias voltage ranging from $400–1000~\mathrm{V}$ through mechanical contact with the ITO surface, with the transducer horn maintained at ground potential throughout. The discharge mechanism was provided by a soft X-ray source (Hamamatsu L12645) directed at the air surrounding the acoustic cavity but away from the microspheres. The entire setup was enclosed in a spacious stainless steel box ($l \times w\times h = 24 \times 23 \times 45.5 \times \mathrm{cm^3}$) to reduce unwanted air currents and prevent X-ray leakage. 
 
 A light source (EL-300 LED Video Light, Jinbei) placed outside the chamber illuminated the particles through transparent windows that provided optical access to the chamber. Particle motion was recorded from the side using a digital camera (DCU224M-GL, Thorlabs), and from the bottom via a $45^\circ$ mirror using a high-speed camera (Phantom VEO410L, Vision Research) operating between $500-1000$ frames per second. Experimental control and data acquisition were managed with a multifunction DAQ device (USB-6002, National Instruments).
 
\subsection*{Simulation}
We perform simulations of stable cluster states using forces calculated via standard time-averaged perturbation theory of the acoustic field~\cite{Bruus.2011all, bruus2012acoustofluidics, silva2014ideal}.
Using first-order perturbation theory and assuming single-frequency driving, the spatial component of the acoustic velocity potential $\phi$ inside the cavity obeys the Helmholtz equation
\begin{equation}
    \nabla^2 \phi(x) = -\frac{\omega^2}{c^2}\phi(x) \,,
\end{equation}
where $\omega$ is the oscillation frequency of the acoustic field and $c$ is the speed of sound in air.

For simplicity, we model our acoustic cavity as a cylinder of radius $R$ and height $h$, with hard-wall boundary conditions on the top and bottom and soft-wall boundary conditions along the side, and we assume that the acoustic velocity potential inside the cavity is given by a resonant mode of the cavity of the form
\begin{equation}
    \phi(r, z, \theta; k_r, k_z, n) = \phi_0 \,J_{n}(k_r r) \, \sin(k_z z) \, \cos(n\theta) \,,
\end{equation}
where $\phi_0$ is a wave amplitude, $J_n(x)$ is the $n$-th order Bessel function (of the first kind), $r, z, \theta$ are cylindrical coordinates specifying a location within the cavity, $k_r, k_z$ are wavevector components in the radial and axial directions respectively, and $n$ is an integer.
We assume that piezo-acoustic driving only excites the fundamental cavity mode characterized by $n=0$, $k_z = \pi / h$, and $k_r = j_{0, 1} / R$, where $j_{n, m}$ is the $m$th zero of the $n$th order Bessel function.
We choose $h=3.4~\mathrm{mm}$ for all simulations.

The scattering of the incoming acoustic field $\phi$ causes particles to be trapped in pressure nodes and induces acoustic scattering interactions between the particles.
In the Rayleigh limit, where particles are much smaller than the wavelength of the acoustic field, the forces on the particles can be computed analytically.
Assuming spherical, dense, and sound-hard particles, the acoustic trap force can be written as the gradient of an effective trap potential~\cite{settnes2012forces, bruus2012acoustofluidics}
\begin{equation}
    U_\mathrm{trap}(\bm{x}) = \pi\rho a^3  \left[\frac{k^2}{3} |\phi(\bm{x})|^2 - \frac{1}{2} |\nabla \phi(\bm{x})|^2\right] \,,
\end{equation}
where $a$ is the particle radius, $\rho$ is the density of air, and $k = \sqrt{k_r^2 + k_z^2} = \omega / c$.
Similarly, the acoustic scattering forces between two particles can also be written as the gradient of an interaction potential~\cite{silva2014ideal, King.2025} that depends on the incoming acoustic field and whose full expression is reproduced in the \textit{SI Appendix}.

Beside these acoustic interactions and the electrostatic interactions due to the particles' charge,we also add a simple soft-sphere potential to minimize particle overlaps, 
\begin{equation}
    U_\mathrm{rep}(\Delta\bm{x}) = \varepsilon \left[1 - \frac{\| \Delta\bm{x}\|}{2a}\right]^\alpha \, \Theta(2a - \| \Delta\bm{x}\|) \,,
\end{equation}
where $\Theta(x)$ is the Heaviside function.
We chose $\varepsilon = 10^5 \,\mathrm{nJ}$ and $\alpha=4$ for all simulations in this paper.
Finally, we assume particles experience Stokes drag, $F_\mathrm{drag} = -6\pi \eta a v$ as they move through the surrounding medium, where $\eta$ is the viscosity of air, and $v$ is the particle velocity.

To identify stable cluster states, we first initialize $n$ particles by randomly drawing their $xy$-positions (but forbidding particle overlaps), and setting their $z$-position to be roughly in the plane of stable levitation, i.e. where the acoustic trap force in the $z$-direction balances gravity (see \textit{SI}).
We further assume that the particles' initial $xy$-velocities are given by a normal distribution with zero mean and standard deviation $v_0$.
To minimize the chance of particles `dropping' out of the trap, we set their initial $z$-velocities to zero.
We then simulate the particles' deterministic motion using MD, and stop the simulations once the damping due to drag forces has caused particles to come to rest in a stable configuration, so that all particle velocities are smaller than a threshold velocity; we choose $v_\mathrm{thresh} = 10 \, \mathrm{\mu m / s}$.
Alternatively, if any particle drops out of the trap, we restart the simulation with newly drawn initial conditions until a stable configuration is obtained.

In total, for $\mathrm{n=2}$ to $\mathrm{n=5}$ particles, we perform 100 simulations for each combination of particle charge $Q \in (0.5 \, \mathrm{pC} , 1.3 \, \mathrm{pC})$, acoustic pressure amplitude $p_0 \in (1800\, \mathrm{Pa}, 2000\, \mathrm{Pa}, 2200\, \mathrm{Pa}, ...,  3600\, \mathrm{Pa})$, initial velocity $v_0 \in (5~\mathrm{mm/s}, 15~\mathrm{mm/s})$, and cavity radius $R \in (10~\mathrm{mm}, 15~\mathrm{mm})$.
These simulations lead to the observed unique cluster states shown in Fig.~\ref{fig2}a (which were obtained using monodisperse particles of radius $a = 0.1375 \, \mathrm{mm}$) and the cluster statistics shown in Figs.~S4-S7 (which were obtained using polydisperse particles, see SI for details).
In addition to the cluster states shown in Fig.~\ref{fig2}a, we also find some stable states that are never observed experimentally.
These additional states contain a long line of particles, which seems to render them unstable in experiments, as discussed in the main text and SI.

\begin{acknowledgments}
We thank D. Kleckner for insightful discussions. We acknowledge the MIBA machine shop at the Institute of Science and Technology Austria for instrumentation support. M.C.H. and C.P.G. acknowledge funding by the Gesellschaft f\"ur Forschungsf\"orderung Nieder\"osterreich under project FTI23-G-011.
\end{acknowledgments}

\input{main.bbl}
\end{document}


\onecolumngrid            
\appendix                 
\section*{Supplementary Information: Electrostatics overcome acoustic collapse \\to assemble, adapt, and activate levitated matter}

\subsection{Methods for charge measurement}

For the precise charge measurement of a particle, as in Fig.~1\textit{D}, we employ a separate but nearly identical acoustic levitation setup specifically equipped with a high-speed side-view camera (Phantom VEO 640L), chamber windows that allow backlighting to the field of view of this camera, an integrated AC electric field, and necessary timing infrastructure on a centralized DAQ. This setup has a slightly larger cavity height of 4.4 mm, compared to the main setup's height of 3.4 mm. Consequently, the electric field in this setup is slightly smaller at the same applied bias compared to the main experimental setup. Hence, we extract the slope of the measured charge as a function of the electric field in this setup ($Q$~\textit{vs.}~$E$ in Fig.~1\textit{D} of the main text), and then rescale this via the ratio of cavity heights to quantitatively estimate the particle charging under the conditions of the main experimental setup. \par

Experiments were performed for 10 voltage values on the bottom plate, from 0 to 900~V. For each voltage, (at most) 60 bounces were performed, discharging before every trial. During each trial, the acoustic field is briefly interrupted to allow the particle (diameter $\simeq288$~\textmu{m}) to bounce once on the bottom substrate before being recaptured by the acoustic trap. To measure the charge buildup during this process, an electric field is applied between the bottom plate and the grounded transducer horn, with a field shape similar to that generated by a parallel--plate capacitor. The electric field $E(t)$ is frequency swept through the resonant frequency of the particle in the acoustic trap ($\approx$~50~Hz), and the resulting trajectory of the sphere is recorded with the high-speed camera. The vertical position of the particle as a function of time $y(t)$ can be obtained through particle tracking and fitted with Newton's second law projected on the vertical direction to extract the charge, as follows:
\begin{equation}
\label{newtonlaw}
\ddot{y} = -g-a\sin2ky-2\beta_{\mathrm{0}}\dot{y}-2\beta_{\mathrm{1}}|\dot{y}|\dot{y}+QE(t)/m
\end{equation}
where the electric field $E(t)$, the acoustic wave number $k$, and the sphere's mass $m$ are known; the vertical velocity $\dot{y}$ and acceleration $\ddot{y}$ can be calculated numerically. The unknowns here are the acoustic amplitude $a$, the linear and quadratic damping coefficients $\beta_0$ and $\beta_1$ from air drag, and charge $Q$ on the sphere, which can be determined by fitting. See Refs.~\cite{kline2020precision, grosjean2023single} for full details.

\subsection{Contribution of electric field to particle launch}
When a particle acquires charge via contact with the biased bottom plate, it also experiences an upward force due to the normal component of the electrostatic force exerted by the biased plate \cite{fish1967conductive}:
\begin{equation}
\label{forceonsphere}
F_e \approx \frac{7}{12}\pi^{3}a^{2}\epsilon_{0}E^2
\end{equation}
When this force exceeds the gravitational force $mg$ ($m$ is the mass of the particle and $g$ is the acceleration due to gravity), the charged sphere can become airborne. The critical electric field for this force balance is:
\begin{equation}
\label{criticalfield}
E_c \approx \sqrt{\frac{16\rho ga}{7\pi^{2}\epsilon_0}}
\end{equation}
where $\rho$ is the density of the particle. Under our experimental conditions, the ITO bias voltage necessary to reach this critical electric field is around 600~V. \par

To ensure the particles acquire charge before levitation, we wait at least 100~ms after applying the bias before activating the acoustic field. At subcritical biases ($\mathrm{<600~V}$), particles on the ITO surface repel prior to lift-off, confirming charge induction. Once levitated, the bias is removed and particles settle into stable levitation within 1~s. At the beginning of the assembly process, particles are propelled upward and oscillate both vertically and horizontally due to the momentum imparted by the acoustic field as well as any residual electric field. Whenever two particles approach within the critical separation $r < r^*$ described in Fig.~1\textit{E}, the short-range acoustic scattering force overcomes Coulomb repulsion, pulling them into a collapsed dimer. As shown in Fig.~1\textit{F, i}, this  dimer is held at the cavity's planar center, where the in-plane restoring force $F_\mathrm{r}$ reaches its maximum. In contrast, spheres that never breach $r^*$ remain separated by long-range Coulomb repulsion; $F_\mathrm{r}$ then confines them laterally, maintaining stable levitation about the cavity center, as shown in Fig.~1\textit{F, ii}.

\subsection{Configuration statistics for n = 2 -- 5}
Here we discuss in detail the statistical results in Fig.~2\textit{C}, which shows the observed occurrence frequency (in percentage) for each configuration, where bar colors indicate stability (teal for stable, yellow for transient). Every bar plot has 100 independent experimental trials, each initiated with particles at rest at random initial positions on the ITO plate biased at 400~V.\par

For $\mathrm{n=2}$, the fully expanded configuration dominates ($>80\%$ occurrence) and is unconditionally stable, whereas the collapsed dimer is transient in most trials, consistent with the instability discussed in the main text. Similarly, fully expanded configurations for $\mathrm{n=3}$ and $\mathrm{n=4}$ also dominate and remain stable once formed. The collapsed configurations (triangle for $\mathrm{n=3}$ and rhombus for $\mathrm{n=4}$) appear much less frequently but also remain stable. The hybrid at $\mathrm{n=3}$, comprising a collapsed dimer and one monomer, occurs slightly more often than the collapsed triangle, but consistently rearranges into a fully expanded triangle and thus is always transient. At $\mathrm{n=4}$, the hybrid configuration composed of a collapsed dimer and two monomers is stable in fewer than half of the trials, whereas the configuration consisting of a collapsed triangle and one monomer remains stable in all observed cases. The double-dimer configuration for $\mathrm{n=4}$ was not observed in this particular dataset, but is always transient when seen otherwise. This configuration has two distinct failure modes: either the dimers combine to form a stable collapsed rhombus, or one of the dimers breaks apart to form one of the other two hybrid configurations for $\mathrm{n=4}$.\par

In contrast to $\mathrm{n=2-4}$, for $\mathrm{n=5}$ the fully expanded configuration no longer dominates---its occurrence is only marginally higher than that of the next most frequent structure. Moreover, two distinct fully expanded structures---the cross and the pentagon---can form. Their relative frequencies are determined by the particles' initial velocities, which in turn depend on the applied ITO bias (Fig.~\ref{1000V}\textit{A}). Counterintuitively, the overall yield of fully expanded configurations is lower at the higher $\mathrm{1000~V}$ ITO bias than at the lower $\mathrm{400~V}$ bias (Fig.~\ref{1000V}\textit{B}). At $\mathrm{1000~V}$ ITO bias ($\sim 1.3~\mathrm{pC}$ of charge per particle)---above the critical voltage for charged particles to become airborne---particles gain enough kinetic energy to bounce repeatedly between the transducer horn and the ITO plate before final assembly, increasing collision rates among particles and thus reducing the probability of forming a fully expanded configuration. \par

Among hybrid configurations for $\mathrm{n=5}$, the one consisting of a dimer plus three monomers is stable more than half of the time; notably, the orientation of the dimer is unpredictable and does not appear to affect stability. The two hybrids  lacking dimers consistently exhibit stability, whereas the structure with two separate collapsed dimers and one monomer is invariably transient. The structure containing a collapsed triangle plus a collapsed dimer was also not observed in this dataset but has been transient in all other observations (Fig.~\ref{1000V}\textit{C}). The collapsed trapezoid for $\mathrm{n=5}$ occurs infrequently but remains consistently stable, as expected.\par

We also performed simulations to further investigate the cluster statistics, following the protocol outlined in the \textit{Methods} section in the main text: for every combination of acoustic pressure and particle charge, we perform 100 simulation runs to measure cluster probabilities.
We assume particles have an average radius of $a = 0.1375 \, \mathrm{mm}$ with a size spread of 10\%.
Note that the charge on each particle is proportional to its surface area; in Fig.~\ref{23pstat}-\ref{5pstat}, we report the charge $Q$ acquired by a particle of radius $a=0.145\, \mathrm{mm}$, which corresponds to the particle size at which the charge measurements reported in Fig.1\textit{D} were performed.

The simulated cluster statistics exhibit a strong dependence on the particle charge and the properties of the cavity.
Fig.~\ref{23pstat} and Fig.~\ref{45pstat} show cluster statistics for $\mathrm{n=2-5}$ particles for various values of the acoustic pressure amplitude $p_0$ at initial velocity $v_0=15~\mathrm{mm/s}$ and particle charge $Q=0.5~\mathrm{pC}$. We also vary the radius $R$ of the cavity, which affects the stiffness of the radial acoustic trap force.
While simulations identify all of the stable cluster ground states that are observed in experiments, the agreement between experimental and simulated cluster statistics remains qualitative at best.
Interestingly, we find that a larger cavity radius of $R=15~\mathrm{mm}$ leads to somewhat better agreement, suggesting that the experimental cavity (composed of a flat-bottom transducer with a radius of 10~mm and a 50~mm by 50~mm square bottom plate) is less stiff than the idealized cylindrical cavity used in simulations. This is reasonable, since the bottom plate of the experimental trap is much larger than the transducer, and simulations assume perfectly absorbing boundary conditions at the cavity boundary.

For higher particle charge of $Q=1.3~\mathrm{pC}$, simulations always result in fully expanded configurations at low pressure values (below 2200~Pa). A possible explanation is that the theoretical calculations of the acoustic scattering force (which are only strictly valid in the Rayleigh limit) underestimate acoustic attractions if particles are close to contact, such that electrostatic repulsion can succeed in forcing the particles apart. At stronger acoustic fields, hybrid structures start to occur at around $2800~\mathrm{Pa}$ for $\mathrm{n=2-5}$; we show a pressure scan for $\mathrm{n=5}$ in Fig.~\ref{5pstat}.

The discrepancy between experimental and simulated cluster statistics suggests that factors not accounted for in our idealized simulations---such as nonreciprocity due to nonuniform particle attributes~\cite{king2025scattered}, higher-order corrections of the scattering and trap forces, velocity-dependent damping from resonance frequency shifts~\cite{andrade2019experimental}, and/or the precise launch conditions of the particles---can significantly influence the statistics of cluster formation.
Furthermore, it is important to note that we do not observe the dimer instability in our simulations, which leads to higher counts of clusters containing dimers compared with experiments.

\subsection{Additional simulation details}
For completeness, we reproduce here the full expression for the acoustic interaction force between two dense and sound-hard particles in the Rayleigh limit.

Consider two particles at locations $\bm{x}_1$ and $\bm{x}_2$ respectively interacting via acoustic scattering.
Following Ref.~\cite{Silva.2014}, the force felt by particle 1 can be written as the gradient of an interaction potential:
\begin{equation}
    U(\bm{x}_1, \bm{x}_2) = \pi \rho a_1^3  k^2 \, \mathrm{Re}\left[\frac{2}{3} \phi^\star(\bm{x}_1)  \phi_\mathrm{sc}(\bm{x}_1, \bm{x}_2) - \frac{1}{k^2} \frac{\partial \phi^\star(\bm{x}_1)}{\partial \bm{x}_1} \cdot \frac{\partial \phi_\mathrm{sc}(\bm{x}_1, \bm{x}_2)}{\partial \bm{x}_1} \right] \,,
\end{equation}
where $\phi(\bm{x})$ is the incoming field given by a resonant mode of the acoustic cavity (as discussed in \textit{Methods}) and $\phi_\mathrm{sc}(\bm{x}_1, \bm{x}_2)$ is the scattering field caused by the presence of particle 2 at $\bm{x}_2$ and felt by particle 1 at $\bm{x}_1$:
\begin{equation}
    \phi_\mathrm{sc}(\bm{x}_1, \bm{x}_2) = -\frac{a_2^3 \, e^{ik \|\Delta \bm{x}\|}}{\|\Delta \bm{x}\|} \, 
    \left[\frac{k^2}{3} \phi(\bm{x}_2) + \frac{ik - \|\Delta \bm{x}\|^{-1}}{2\|\Delta \bm{x}\|} \Delta \bm{x} \cdot \frac{\partial \phi(\bm{x}_2)}{\partial \bm{x}_2} \right] \,,
\end{equation}
where $\Delta \bm{x} = \bm{x}_1 - \bm{x}_2$. The forces on the particles are then given by $F_1 = -\partial_{\bm{x}_1} U(\bm{x}_1, \bm{x}_2)$ and $F_2 = -\partial_{\bm{x}_2} U(\bm{x}_2, \bm{x}_1)$.
Note that the forces are not necessarily reciprocal, especially if the incoming acoustic field is not homogeneous.

As mentioned in \textit{Methods}, during simulations we initialize the particles roughly in the plane of stable levitation, where acoustic trap and gravity forces balance.
Assuming for just this section that the cavity wave is a simple standing wave with no $r$-dependence, the location of the levitation plane can be estimated to be
\begin{equation}
    z^\star \approx -\frac{4\,\rho_\mathrm{p} \, \rho_\mathrm{m} \, g \, c^2}{5 \, p_0^2 \, k^2}\,,
\end{equation}
where $\rho_\mathrm{p}$ is the density of the particles, $\rho_\mathrm{m}$ is the density of the medium (air), $g$ is the gravity on Earth, $c$ is the speed of sound in air, and $p_0 = ck \rho_\mathrm{m}\phi_0$ is the pressure amplitude of the incoming acoustic wave.

\subsection{Non-reciprocal and many-body interactions}
As mentioned in the main text, we observe several phenomena that may be signatures of non-reciprocal and/or many-body effects in our system. Before discussing these, we briefly introduce why they should be expected to occur.  

While our simulations do take into account non-reciprocity due to an inhomogeneous incoming cavity wave, they do not capture higher-order effects due to nonuniform particle sizes, densities, or compressibilities.

Ref.~\cite{king2025scattered} recently showed that, even for a system that \textit{only} breaks the assumption of identical size, the force between two particles $i$ and $j$ has the form
\begin{equation}
\hat{\mathbf{F}}_{ij}(\mathbf{r}_{ij}) = (1 + \hat{\chi}_{ij}) \mathbf{F}^K_{ij}(\mathbf{r}_{ij}).
\label{eq:2}
\end{equation}
If the incoming cavity is a simple standing wave with no $r$-dependence, $\mathbf{F}^K_{ij}(\mathbf{r}_{ij})$ is the reciprocal, second-order expansion of the scattering force, which reduces to Eq.~(2) if $a\ll r\ll\lambda$, but more generally is given by the so-called ``K\"onig'' expression, as shown in Ref.~\cite{king2025scattered}.
The term $\hat{\chi}_{ij}$ is an interaction matrix that depends on the particle radii, and encodes the non-reciprocity in that $\hat{\chi}_{ij} \ne \hat{\chi}_{ji}$.  In the case that the objects are not spherical, it can be expected that this matrix additionally depends on the orientation of each object.
For sound-hard and incompressible particles, the non-reciprocal term scales as $\hat\chi_{ij} \propto (ka)^2$.
In our system, where $(ka)^2 \approx 0.06$, hence leading to minor but not entirely negligible corrections to the second-order scattering results used in simulations (see the previous section). It is therefore plausible that these non-reciprocal forces could lead to dynamic effects.

Ref.~\cite{st2023dynamics} recently performed simulation studies to investigate the many-body aspects of the scattering force. These exist due to ``multi-particle scattering'', \textit{i.e.}, sound scattering from one particle off of another can affect field amplitudes at a third.  These interactions are much harder to approach analytically, and to the best of our knowledge so far are only treated in simulations.  There, it has been shown that even with systems composed of identical spheres, effective non-conservative forces can be observed due to asymmetries in particle configurations. In our system, many-body interactions are less likely to contribute to observed behavior than non-reciprocal ones due to the fact that they are a ``third order'' scattering effect, though we cannot strictly exclude them.  

A potential manifestation of non-reciprocity in our experimental system is the frequently observed dimer instability. Dimers are composed of two individual particles acting effectively as a single composite scatterer, which has a particularly asymmetric configuration from a sound-scattering perspective. Due to this inherent asymmetry, even minor rotations or shifts can alter the scattering profile of dimers, leading to pronounced orientation dependence. In clusters beyond dimers (\textit{i.e.}, clusters with more than two constituent particles), the configurations become less asymmetric, hence reducing their orientation sensitivity. Moreover, according to Ref.~\cite{lim2022mechanical}, the acoustic binding energy scales with the cluster size, stabilizing larger clusters with increased total adhesive energy. Therefore, the observed dimer instability is quite conceivably connected to non-reciprocity as described by King \textit{et al}.~\cite{king2025scattered}.

Another example of non-reciprocal interactions in our system can be seen in the ``expanded chaser dynamics'' in Movie S14. The initially stable expanded triangle configuration becomes unstable due to the rapid rotational motion of one of the three particles, which is visibly aspherical and hence has an anisotropic scattering profile. As the particle continuously reorients itself due to torque from the acoustic field, these anisotropic forces dynamically modify the scattering forces acting on other particles, resulting in the observed ``chasing'' behavior \cite{lim2024acoustic}. Thus, this scenario is highly likely a direct demonstration of non-reciprocal acoustic interactions that arise from particle asymmetry and anisotropic scattering.

\setcounter{figure}{0}
\renewcommand{\thefigure}{S\arabic{figure}}

\begin{figure}[p]
\centering
\includegraphics[width=0.6\textwidth]{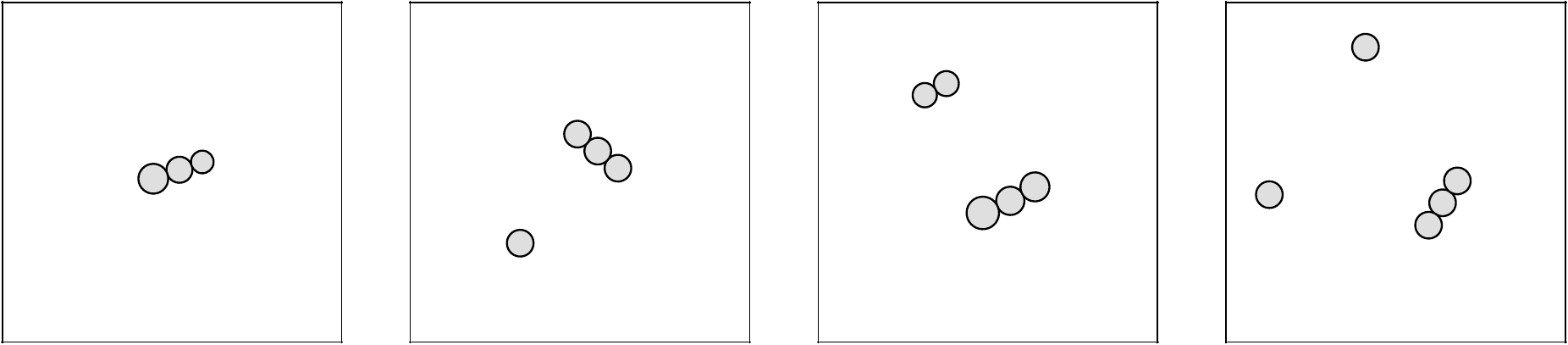}
\caption{Simulated configurations for $\mathrm{n=3-5}$ particles containing chains longer than 2 particles. These simulated configurations do not appear as long-lived stable states in experiments due to perturbations.}
\label{chains}
\end{figure}
\clearpage

\begin{figure}[p]
\centering
\includegraphics[width=\textwidth]{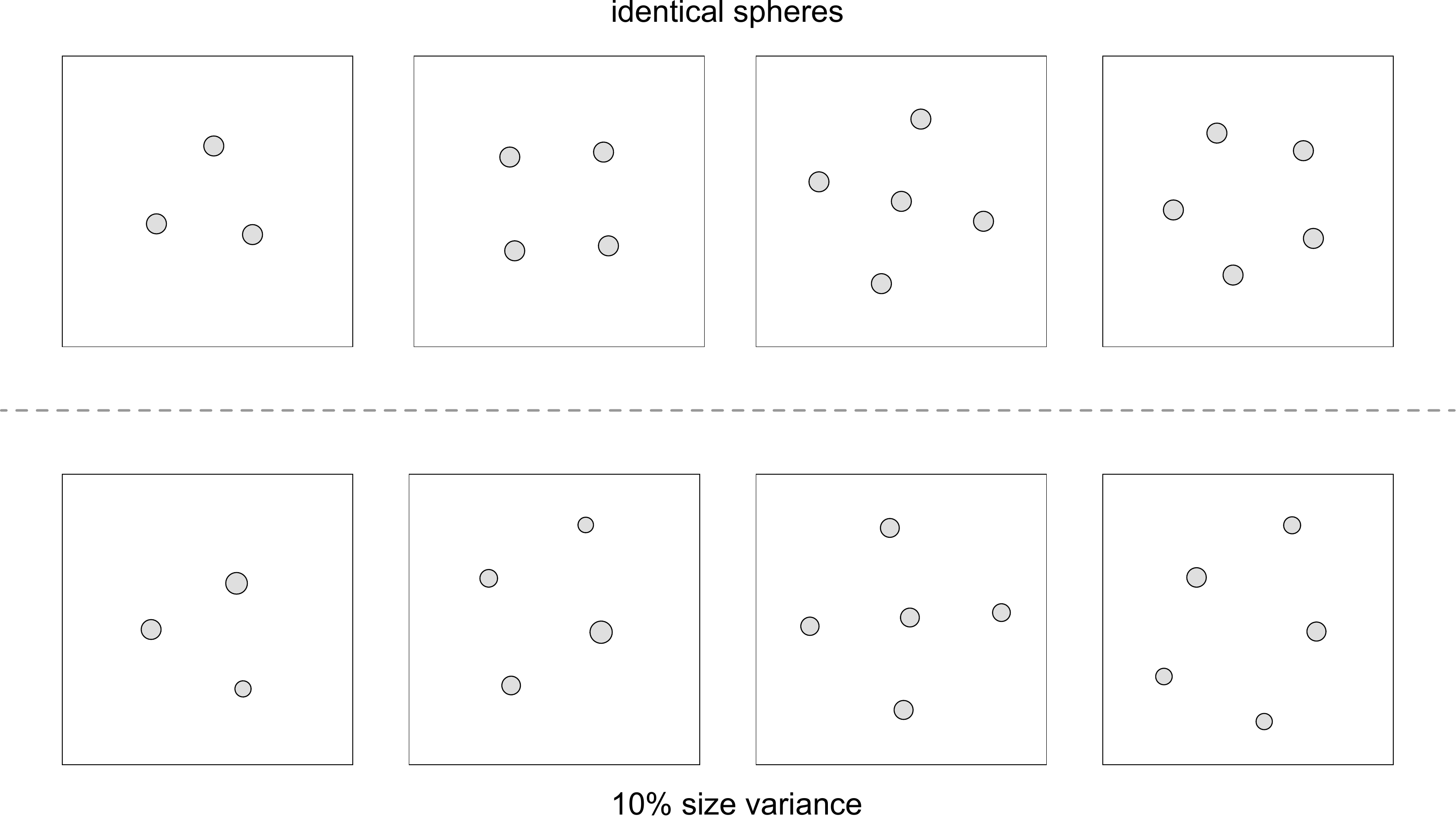}
\caption{Simulated fully expanded configurations for $\mathrm{n=3-5}$ identical particles (top) and particles with $10\%$ size variance (bottom), corresponding to the experimental particle size distribution. The comparison illustrates the impact of slight polydispersity on structural symmetry.}
\label{polydispersity}
\end{figure}
\clearpage

\begin{figure}[p]
\centering
\includegraphics[width=150mm]{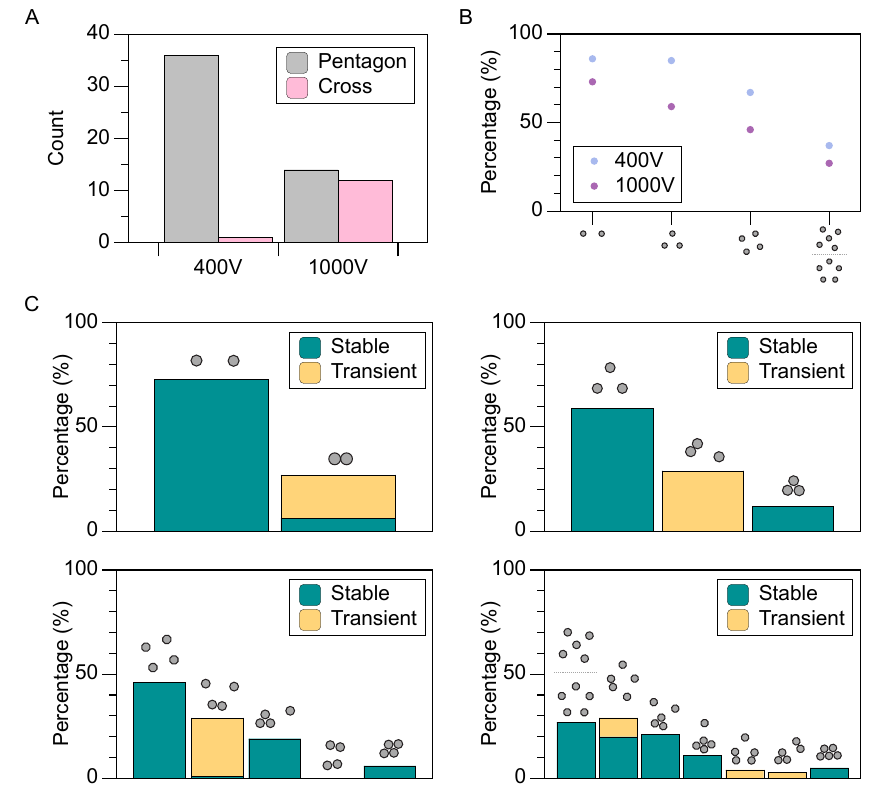}
\caption{\textbf{Effect of ITO bias voltage on structure formation and degeneracy.} (A) Absolute counts of the two distinct $\mathrm{n=5}$ configurations---pentagon (gray) and cross (pink)---observed at both 400~V and 1000~V ITO bias voltages. (B) Percentage of fully expanded assemblies among all observed structures for particle counts $\mathrm{n=2-5}$. Light blue circles show data from trials with ITO plate biased at 400~V and dark purple circles show data at 1000~V. (c) Structure formation statistics for $\mathrm{n=2-5}$ particles with ITO plate biased at 1000~V. Experimental bar plots show the occurrence frequency and stability of each structure, expressed as a percentage. Stable occurrences (teal) persist for more than 15 seconds, while transient occurrences (yellow) rearrange or disintegrate within 15 seconds.}
\label{1000V}
\end{figure}
\clearpage

\begin{figure}[p]
\centering
\includegraphics[width=0.9\textwidth]{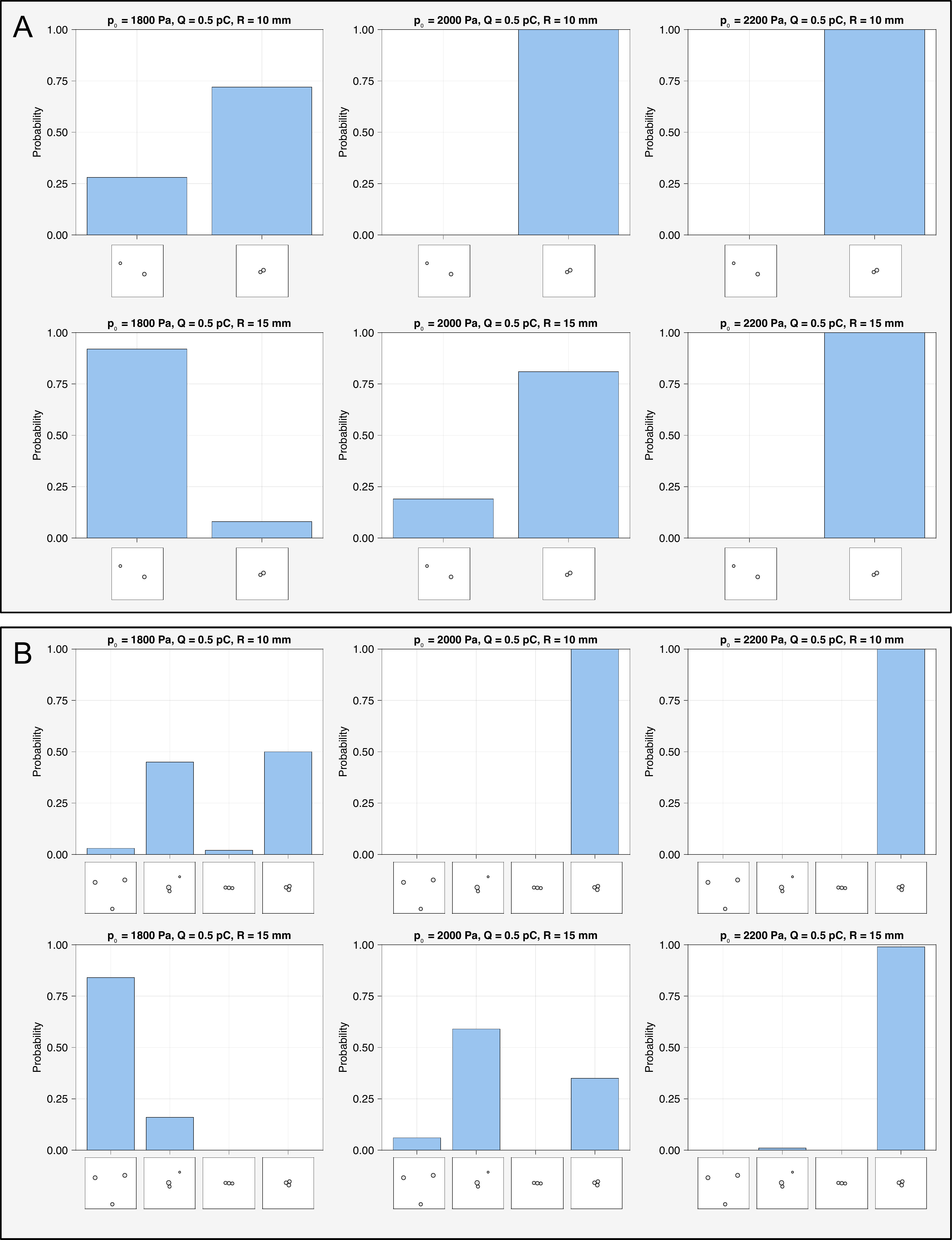}
\caption{Simulated cluster statistics for particles with 10\% size variance for different pressure amplitude $p_0$ and trap radius $R$ at mean particle charge $Q=0.5~\mathrm{pC}$. (A) shows simulations $\mathrm{n}=2$ particles, (B) shows simulations with $\mathrm{n}=3$ particles.}
\label{23pstat}
\end{figure}
\clearpage

\begin{figure}[p]
\centering
\includegraphics[width=0.9\textwidth]{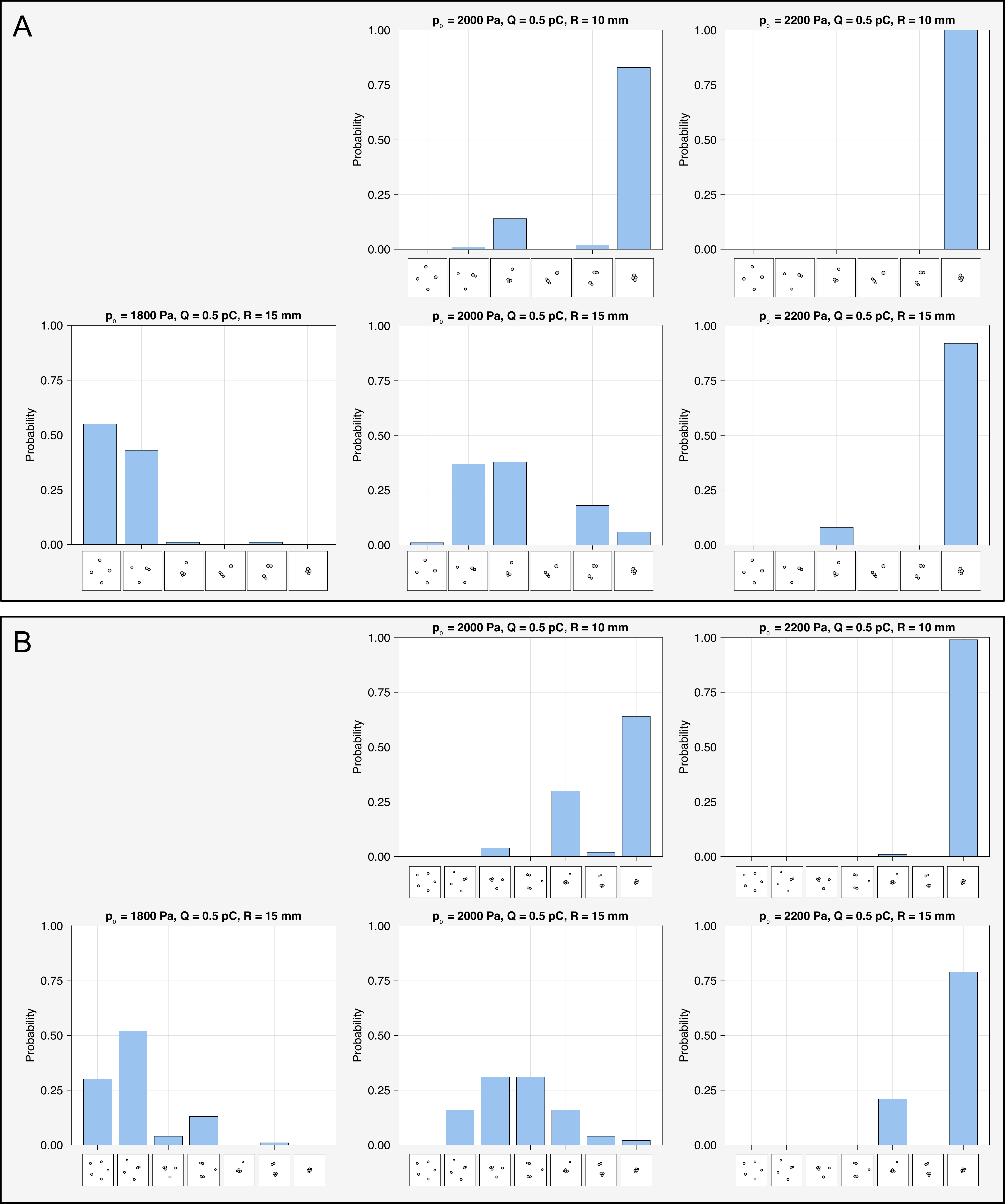}
\caption{Simulated cluster statistics for particles with 10\% size variance for different pressure amplitude $p_0$ and trap radius $R$ at mean particle charge $Q=0.5\,\mathrm{pC}$.  (A) shows simulations $\mathrm{n}=4$ particles, (B) shows simulations with $\mathrm{n}=5$ particles. The bar corresponding to the fully expanded structure in (B) counts both possible fully expanded states (the pentagon and the cross). Note that there is no data for $R = 10\,\mathrm{mm}$, and $p_0 = 1800\,\mathrm{Pa}$; at these parameter values, particles drop out of the trap so often that no statistics could be collected.}
\label{45pstat}
\end{figure}
\clearpage

\begin{figure}[p]
\centering
\includegraphics[width=0.9\textwidth]{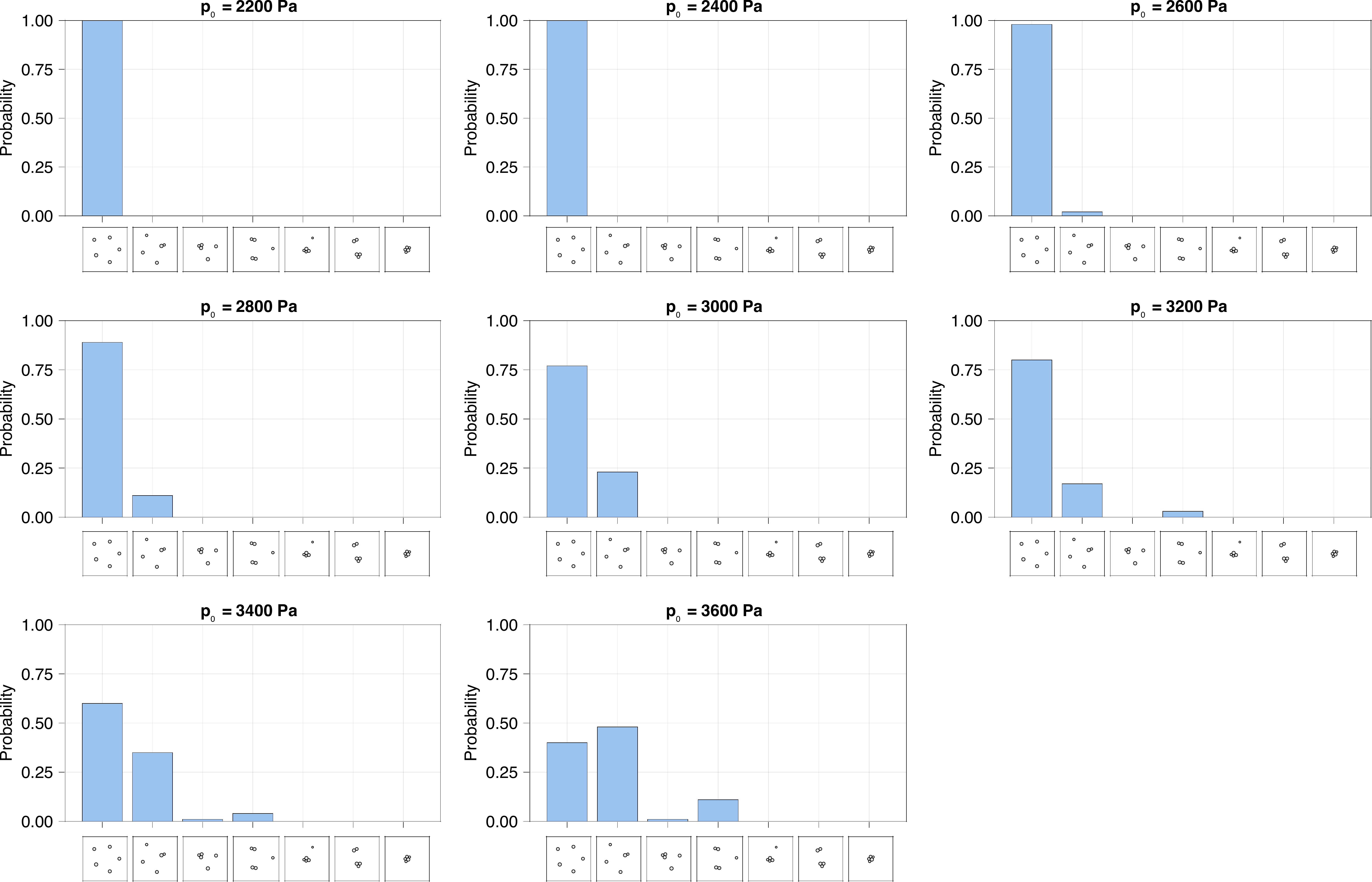}
\caption{Simulated cluster statistics of particles with 10\% size variance for different pressure amplitude $p_0$ at initial velocity $v_0=15\,\mathrm{mm/s}$, charge $Q = 1.3\,\mathrm{pC}$ and trap radius $R=10\,\mathrm{mm}$.}
\label{5pstat}
\end{figure}
\clearpage

\clearpage                
\subsection*{SI Movies}

\movie{Bottom view of two uncharged particles ($\mathrm{250-300~\mu m}$ in diameter) levitating and forming a collapsed dimer due to attractive acoustic scattering forces. Playback speed is real time. Scale bar corresponds to $\mathrm{1~mm}$.}

\movie{Bottom view of two like-charged particles ($\mathrm{250-300~\mu m}$ in diameter) levitating off of the ITO plate biased at 500~V and forming an expanded dimer due to repulsive Coulomb forces. Playback speed is real time. Scale bar corresponds to $\mathrm{1~mm}$.}

\movie{Bottom view of four particles ($\mathrm{250-300~\mu m}$ in diameter) forming a transiently stable hybrid structure composed of two separate monomers and a collapsed dimer. Once the unstable dimer breaks apart, the two constituent particles remain separated due to Coulomb repulsion and form a stable fully expanded square-like structure with the other two monomer particles. Playback speed is real time. Scale bar corresponds to $\mathrm{1~mm}$.}

\movie{Bottom view of a compact dimer consisting of two uncharged particles ($\mathrm{250-300~\mu m}$ in diameter), which spontaneously begin to oscillate in the acoustic field. The oscillations amplify over time and eventually the two constituent particles temporarily escape their bond. After the particles return to the collapsed dimer configuration due to attractive scattering forces, the spontaneous destabilizing oscillations can be seen again. Playback speed is real time. Scale bar corresponds to $\mathrm{1~mm}$.}

\movie{Side view of the process shown in Movie S4. Playback speed is real time.}

\movie{Bottom view of four like-charged, levitated particles undergoing a single-step discharge process. Particles are $\mathrm{250-300~\mu m}$ in diameter, and applied bias to the ITO plate is $\mathrm{500~V}$. Playback speed is real time. X-ray emission starts at $\mathrm{t=0~s}$ and stops at $\mathrm{t=10~s}$. The fully expanded square becomes smaller during X-ray exposure, and two of the four particles form a collapsed dimer at $\mathrm{t=10~s}$. The unstable collapsed dimer temporarily breaks apart five times within 2 seconds, before forming a stable collapsed triangle with one of the other two particles. The final structure consists of the collapsed triangle with one separated monomer, and remains stable in the absence of further X-ray emission. Scale bar corresponds to $\mathrm{1~mm}$.}

\movie{Bottom view of three like-charged, levitated particles undergoing a two-step discharge process. Particles are $\mathrm{250-300~\mu m}$ in diameter, and applied bias to the ITO plate is $\mathrm{500~V}$. Playback speed is real time. X-ray emission starts at $\mathrm{t=0~s}$ and pauses at $\mathrm{t=5~s}$, then resumes at $\mathrm{t=10~s}$ and stops at $\mathrm{t=15~s}$. The fully expanded triangle becomes smaller during the first X-ray exposure, and snaps into a collapsed triangle during the second X-ray exposure. Scale bar corresponds to $\mathrm{1~mm}$.}

\movie{Bottom view demonstrating the reconfiguration process of five like-charged particles via the bounce-charging method. Particles are $\mathrm{250-300~\mu m}$ in diameter, and applied bias to the ITO plate is $\mathrm{500~V}$. Playback speed is real time. The initial structure (collapsed rhombus plus one monomer) is formed by levitating particles initially in contact with the bottom ITO plate biased at $\mathrm{500~V}$. During each bounce, the acoustic field is interrupted for $\mathrm{70~ms}$ to allow the particles to bounce once on the bottom ITO plate, which is maintained at $\mathrm{500~V}$ only during the bounce. After seven bounces, the assembly returns to its initial configuration. Scale bar corresponds to $\mathrm{1~mm}$.}

\movie{Side view of the process shown in Movie S3. Playback speed is real time.}

\movie{Bottom view of the ``acoustic clock'' assembly. Playback speed is real time. Particles are $\mathrm{180-212~\mu m}$ in diameter, and applied bias to the ITO plate is $\mathrm{500~V}$. Upon levitation, the charged particles quickly assemble into a large, semi-expanded structure, consisting of a collapsed core with 10 particles surrounded by an outer ring of 6 particles. The core spontaneously starts to rotate clockwise once formed, resembling the dial of a clock. Scale bar corresponds to $\mathrm{1~mm}$.}

\movie{Bottom view of the ``Ferris wheel'' assembly. Playback speed is 1/10 of the real time. Particles are $\mathrm{180-212~\mu m}$ in diameter, and applied bias to the ITO plate is $\mathrm{600~V}$. The collapsed core rotates counterclockwise, inducing an edge wave on the outer ring of satellite particles due to eletrostatic interactions. Scale bar corresponds to $\mathrm{1~mm}$.}

\movie{Bottom view of the ``magic ring'' assembly. Playback speed is 1/2 of the real time. Particles are $\mathrm{180-212~\mu m}$ in diameter, and applied bias to the ITO plate is $\mathrm{1200~V}$. A collapsed cluster of seven particles resembles the ``diamond'' on a ring. Once formed, the diamond rotates, causing the other particles to oscillate. The entire structure undergoes transitions between a breathing mode, where the ring expands and contracts, and an edge mode, where particles oscillate one after another. Eventually, the diamond's rotation slows down and the structure becomes relatively still. Scale bar corresponds to 1~mm.}

\movie{Bottom view showing the size tunability of a large semi-expanded structure. Playback speed is real time. Particles are $\mathrm{250-300~\mu m}$ in diameter, and applied bias to the ITO plate is $\mathrm{1000~V}$. The structure comprises a stationary, collapsed triangular core surrounded by an outer ring that contracts when the piezo‐driver voltage (before amplification) is raised by $\mathrm{0.5~V_{rms}}$. Scale bar corresponds to $\mathrm{1~mm}$. }

\movie{Bottom view of the rotating Coulomb triangle. Playback speed is 1/2 of the real time. Particles are $\mathrm{180-212~\mu m}$ in diameter, and applied bias to the ITO plate is $\mathrm{600~V}$. Three like-charged particles levitate to form a fully expanded triangle. One of the particles is not perfectly spherical, and constantly rotates around its own preferred axis. At t = 19~s, the self-rotation of the non-spherical particles triggers rotation of the triangle.}

\input{si.bbl}

%% file: main.bbl
\providecommand{\noopsort}[1]{}\providecommand{\singleletter}[1]{#1}%